\documentclass[twocolumn]{aa} % for a paper on 1 column  

\usepackage{natbib}
\usepackage{amsmath}
\usepackage{graphicx}
\usepackage{txfonts}
\usepackage{textcomp}

\makeatletter
\def\@biblabel#1{}
\makeatother
\newcommand{\mjup}{M$_{\rm J}$\,}
\newcommand{\msun}{M$_\odot$\,}

\newcommand{\ms}{m\,s$^{-1}$\,}
\newcommand{\teff}{T$_{{\rm eff}}$\,}
\renewcommand{\cite}{\citealp}

\begin{document}

\title{A planetary system and a highly eccentric brown dwarf around the giant stars HIP\,67851 and HIP\,97233.
\thanks{Based on observations collected at La Silla - Paranal Observatory under
programs ID's 085.C-0557, 087.C.0476, 089.C-0524, 090.C-0345 and through the Chilean Telescope Time under programs
ID's CN 12A-073, CN 12B-047 and CN 13A-111.}}

  \titlerunning{}
   \author{M. I. Jones \inst{1}
           \and J. S. Jenkins \inst{2}
           \and P. Rojo \inst{2}
           \and C. H. F. Melo \inst{3}
           \and P. Bluhm \inst{4}}
         \institute{Department of Electrical Engineering and Center of Astro-Engineering UC, Pontificia Universidad
         Cat\'olica de Chile, Av. Vicuña Mackenna 4860, 782-0436 Macul, Santiago, Chile \\\email{mjones@aiuc.puc.cl}
         \and Departamento de Astronom\'ia, Universidad de Chile, Camino El Observatorio 1515, Las Condes, Santiago, Chile 
         \and European Southern Observatory, Casilla 19001, Santiago, Chile
         \and Departamento de Astronom\'ia, Universidad de Concepción, Casilla 160-C, Concepción, Chile}

   \date{}

% \abstract{}{}{}{}{} 
% 5 {} token are mandatory
 
  \abstract
{So far more than 60 substellar companions have been discovered around giant stars. These systems present physical and orbital
properties that contrast to those detected orbiting less evolved stars. }
{We are conducting a radial velocity survey of 166 bright giant stars in the southern hemisphere. The main goals of our project
are to detect and characterize planets in close-in orbits around giant stars in order to study the effects of the host
star evolution on their orbital and physical properties.}
{We have obtained precision radial velocities for the giant stars HIP\,67851 and HIP\,97233 that have revealed 
periodic signals, which are most likely induced by the presence of substellar companions.}
{We present the discovery of a planetary system and an eccentric brown dwarf orbiting the giant stars HIP\,67851 and HIP\,97233, 
respectively. The inner planet around HIP\,67851 has a period of 88.8 days, a projected mass of 1.4 \mjup and an eccentricity of 0.09. 
After Kepler 91\,{\it b}, HIP\,67851\,{\it b} is the closest-in known planet orbiting a giant star. Although the orbit of the outer object is not fully constrained, it is 
likely a super-Jupiter. The brown dwarf around HIP\,97233 has an orbital period of 1058.8 days, a minimum mass of 20.0 \mjup and an 
eccentricity of 0.61. This is the most eccentric known brown dwarf around a giant star.}
   {}

   \keywords{techniques: radial velocities - Planet-star interactions - (stars:) brown dwarfs}

   \maketitle
%
%________________________________________________________________

\section{Introduction}

Among the currently known ensemble of exoplanets, only a small fraction have been detected around giant stars, mainly because most of the radial velocity (RV) surveys have 
targeted solar-type stars thus far. However, such planets have properties that are very different than those orbiting main-sequence stars. 
In particular, giant planets are more common ($\sim$ 5 \% for MS host stars versus $\sim$ 10 \% for giant host stars; D\"{o}llinger et al. 
\cite{DOL09}), they are not found in close-in orbits ($a$ $\lesssim$ 0.6 AU; 
Johnson et al. \cite{JOH07}), and they are significantly more massive than planets orbiting lower-mass stars (Jones et al. \cite{JON14}).  
Additionally, the fraction of brown dwarfs (BD) in relatively close-in orbits around giant 
stars also seems to be higher compared to less evolved host stars (Mitchell et al. \cite{MIT13}).  
These peculiar properties of exoplanets around giant stars give us information regarding the role of the stellar mass in the formation
of substellar objects and the effect of the stellar evolution in the orbital and physical properties of these systems (see discussion in Jones et
al. \cite{JON14}). \newline \indent
In 2009 we began the EXPRESS project ({\bf EX}o{\bf P}lanets a{\bf R}ound {\bf E}volved {\bf S}tar{\bf S}; Jones et al. \cite{JON11}), aimed at detecting 
planets around evolved (giant) stars via precision radial velocities, and to study their orbital properties. So far, we have discovered two 
planetary systems (Jones et al. \cite{JON13,JON14}) plus the three substellar objects presented in this paper. 
Interestingly, three of these are super-Jupiters and two are in relatively close-in orbits ($a \lesssim$ 0.6 AU). \newline \indent
In this paper we present a planetary system discovered orbiting the star HIP\,67851, along with the discovery of a highly eccentric brown dwarf orbiting at 
$\sim$ 2.5 AU from the star HIP\,97233. The inner planet orbiting HIP\,67851 is located at an orbital distance of 0.46 AU whereas the outer object has a 
much larger separation ($a \gtrsim$ 4 AU).
In Sect. 2 we describe the observations and the data reduction. In Sect. 3 we briefly present the radial velocity computation method. In Sect. 4 we summarize
the main properties of the host stars. In Sect. 5 we present the orbital parameters of HIP67851\,{\it b,\,c} and HIP97233\,{\it b}. Finally, in Sect. 6 we
present the conclusions.

\section{Observations and data reduction}

During the past five years we have collected $\sim$ 15 spectra of each of 166 targets in the EXPRESS program 
(Jones et al. \cite{JON11,JON13}), using three different high resolution optical spectrographs installed in Chile.
Among these targets, we have followed-up all of them showing RV variations above the $\sim$ 30 \ms level, to determine 
whether they exhibit periodic variations that might be induced by orbiting planets. 
In particular, we took a total of 60 spectra for HIP\,67851 and 41 spectra for HIP\,97233, using the Fiber-fed Extended Range Optical Spectrograph 
(FEROS; Kaufer et al. \cite{KAU99}) and CHIRON (Tokovinin et al. \cite{TOK13}). 
Since these two stars are relatively bright, we obtained a signal-to-noise $\gtrsim$ 100 with relatively short exposure times ($\sim$ 5 and 15 minutes
for FEROS and CHIRON, respectively).
The FEROS spectra were extracted using the FEROS data reduction system, while for CHIRON we
used the reduction pipeline that is provided for the users. These two pipelines perform a standard reduction of echelle spectra (bias subtraction, 
flat-fielding correction, order tracing, extraction and wavelength calibration). For more details see Jones et al. (\cite{JON13,JON14}).

\section{Radial velocity computations \label{sec_RVC}}

The method used to extract the radial velocities has been explained in details in previous papers (Jones et al. 
\cite{JON13,JON14}).
Since we have used different instruments, we had to apply two different methods, namely, the simultaneous calibration method 
(Baranne et al. \cite{BAR96}) for FEROS data and the I$_2$ cell technique (Butler et al. \cite{BUT96}) 
for CHIRON spectra. In the former case, we computed the cross-correlation function between the observed spectra and a template, which 
corresponds to a high S/N observation
of the same star. This procedure is repeated independently for $\sim$ 50 \AA\, chunks. 
We corrected the resulting velocity by the nightly drift, which is computed by cross-correlating the simultaneous calibration
spectrum that is recorded in the sky fiber. Finally, we add the barycentric correction, which leads to the total final radial velocity at
each observational epoch. The typical precision that we achieved for our targets is $\sim$ 5-7 \ms (although it is possible to reach $\sim$ 3 \ms 
in bright objects. See Jones et al. \cite{JON13}).  
For the I$_2$ cell observations, we modeled the observed spectrum of the star, which also contains the I$_2$ absorption spectrum, in a similar fashion
as presented in Butler et al. (\cite{BUT96}). However, we use a simpler model for the instrumental point-spread-function, including only three 
Gaussians (a central profile of fixed width and varying height and two flanking profiles with varying height and width) and we 
applied this method to 5 \AA\, chunks. 
Although our model was optimized for FECH spectra, it leads to a precision of $\sim$ 5 \ms for CHIRON, which is well below the RV 
noise introduced by stellar pulsations in giant stars (e.g. Sato et al. \cite{SAT05}).

\begin{table}
\centering
\caption{Stellar parameters of the host stars. \label{stellar_par}}
\begin{tabular}{lccc}
\hline\hline
               & HIP\,67851 & HIP\,97233 \\
\hline
Spectral Type         & K0III             & K0    \\ 
B-V (mag)             & 1.01              & 1.00  \\
V (mag)               & 6.17              & 7.34  \\ 
Parallax (mas)        & 15.16 $\pm$ 0.39  & 9.39  $\pm$ 0.70 \\
%Distance (pc)         & 66.0  $\pm$ 1.7  & 106.5 $\pm$      \\ 
\teff (K)             & 4890  $\pm$ 100   & 5020  $\pm$ 100  \\
%log\,L (L$_\odot$)    & 1.244 $\pm$ 0.046 & 1.204 $\pm$ 0.076\\
L (L$_\odot$)         & 17.55 $\pm$ 2.64  & 16.00 $\pm$ 5.69 \\
log\,g (cm\,s$^{-2}$) & 3.15  $\pm$ 0.20  & 3.26  $\pm$ 0.20 \\
{\rm [Fe/H]} (dex)    & 0.00  $\pm$ 0.10  & 0.29  $\pm$ 0.13 \\
$v$\,sin$i$ (km\,s$^{-1}$)& 1.8   $\pm$ 0.9   & 2.2   $\pm$ 0.9 \\
M$_\star$ (\msun)     & 1.63  $\pm$ 0.22  & 1.93  $\pm$ 0.11 \\
R$_\star$ (R$_\odot$) & 5.92  $\pm$ 0.44  & 5.34  $\pm$ 0.55 \\
\hline\hline
\end{tabular}
\end{table}
\begin{figure}[]
\centering
\includegraphics[width=9cm,height=8cm,angle=0]{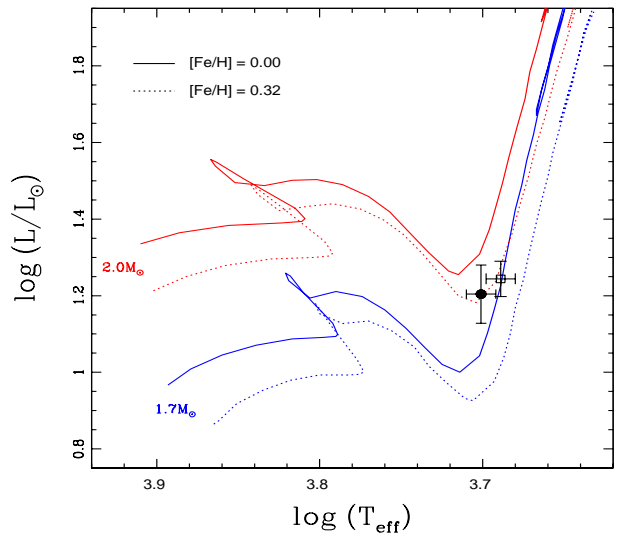}
\caption{Position of HIP\,67851 (open square) and HIP\,97233 (filled circle) in the HR diagram. Different evolutionary tracks (Salasnich et al. 
\cite{SAL00}) for 1.7 \msun and 2.0 \msun are overplotted. The solid and dotted lines correspond to [Fe/H] = 0.0 and [Fe/H] = 0.32, respectively.
\label{HIP67851_HIP97233_position}}
\end{figure}
\begin{figure*}[t!]
\centering
\includegraphics[width=13cm,height=10.5cm,angle=0]{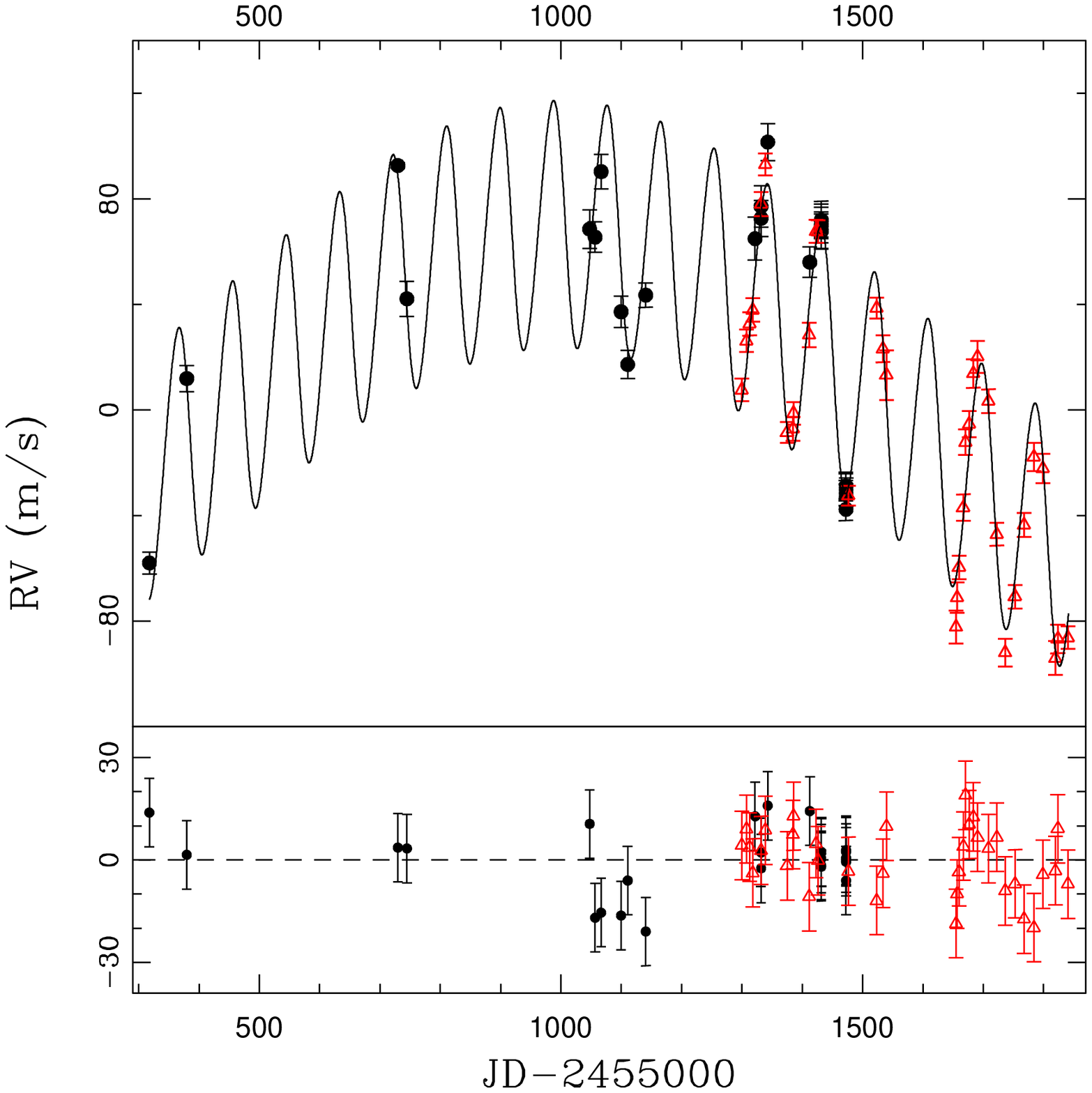}
\caption{Upper panel: Radial velocity curve of HIP\,67851. The red triangles and black dots
correspond to CHIRON and FEROS velocities, respectively. One possible Keplerian solution with two giant planets is overplotted (black solid line).
Lower panel: Residuals from the Keplerian fit. The RMS around the fit is 9.4 \ms.  \label{HIP67851_vels}}
\end{figure*}

\section{Stellar properties}

The main stellar properties of HIP\,67851 and HIP\,97233 are listed in Table \ref{stellar_par}. The colors and parallaxes were taken from the 
Hipparcos catalog (Van Leeuwen \cite{VAN07}), as well as their spectral types. The atmospheric parameters of these two stars 
were computed using the MOOG code (Sneden \cite{SNE73}), by matching the curve of growth 
of $\sim$ 150 Fe {\sc i} and $\sim$ 20 Fe {\sc ii} weak lines, as explained in Jones et al. (\cite{JON11}). 
Figure \ref{HIP67851_HIP97233_position} shows the position of HIP\,67851 (open square) and HIP\,97233 (filled circle) in the HR diagram. 
For comparison, two evolutionary tracks for 1.7 \msun and 2.0 \msun from Salasnich et al. (\cite{SAL00}) models are overplotted, which are the closest 
to the position of these stars. The solid and dotted lines correspond to [Fe/H] = 0.0 and [Fe/H] = 0.32, respectively. 
As can be seen, the two stars are ascending the red giant branch (RGB), although HIP\,97233 is closer to the transition region between the end of 
the subgiant phase and the base of the RGB. Following the procedure described in Jones et al. (\cite{JON11}), we derived a mass of 
1.63 $\pm$ 0.22 \msun and stellar radius of 5.92 $\pm$ 0.44 for HIP\,67851. 
Similarly we obtained a mass of 1.93 $\pm$ 0.11 \msun and radius of 5.34 $\pm$ 0.55 R$_\odot$ for HIP\,97233.
%The masses derived for HIP\,67851 and HIP\,97233 by Jones et al. (\cite{JON11}) are 1.67 $\pm$ 0.1 
%\msun and 1.96 $\pm$ 0.1 \msun, respectively. 
For comparison, we obtained masses for these two stars using the method presented in da Silva et al. 
(\cite{daSIL06})\footnote{http://stev.oapd.inaf.it/cgi-bin/param}. We used the spectroscopic values given in 
Table \ref{stellar_par} and we corrected the V magnitudes with the A$_V$ values obtained from Arenou et al. (\cite{ARE92}) extinction maps. 
In addition, we used the Girardi et al. (\cite{GIR02}) models. We obtained a mass of 1.41 $\pm$ 0.18 \msun and radius of 5.95 $\pm$ 0.36 R$_\odot$ for 
HIP\,67851 and a mass of 1.84 $\pm$ 0.14 \msun and radius of 5.20 $\pm$ 0.50 R$_\odot$ for HIP\,97233. In both cases the mass and radius derived by the two 
methods agree within one sigma. 
%Further details are found in Jones et al. (\cite{JON11}).

\section{Orbital parameters}

\subsection{HIP\,67851\,{\it b,\,c}}

\begin{figure}[h]
\centering
\includegraphics[width=7.0cm,height=8cm,angle=0]{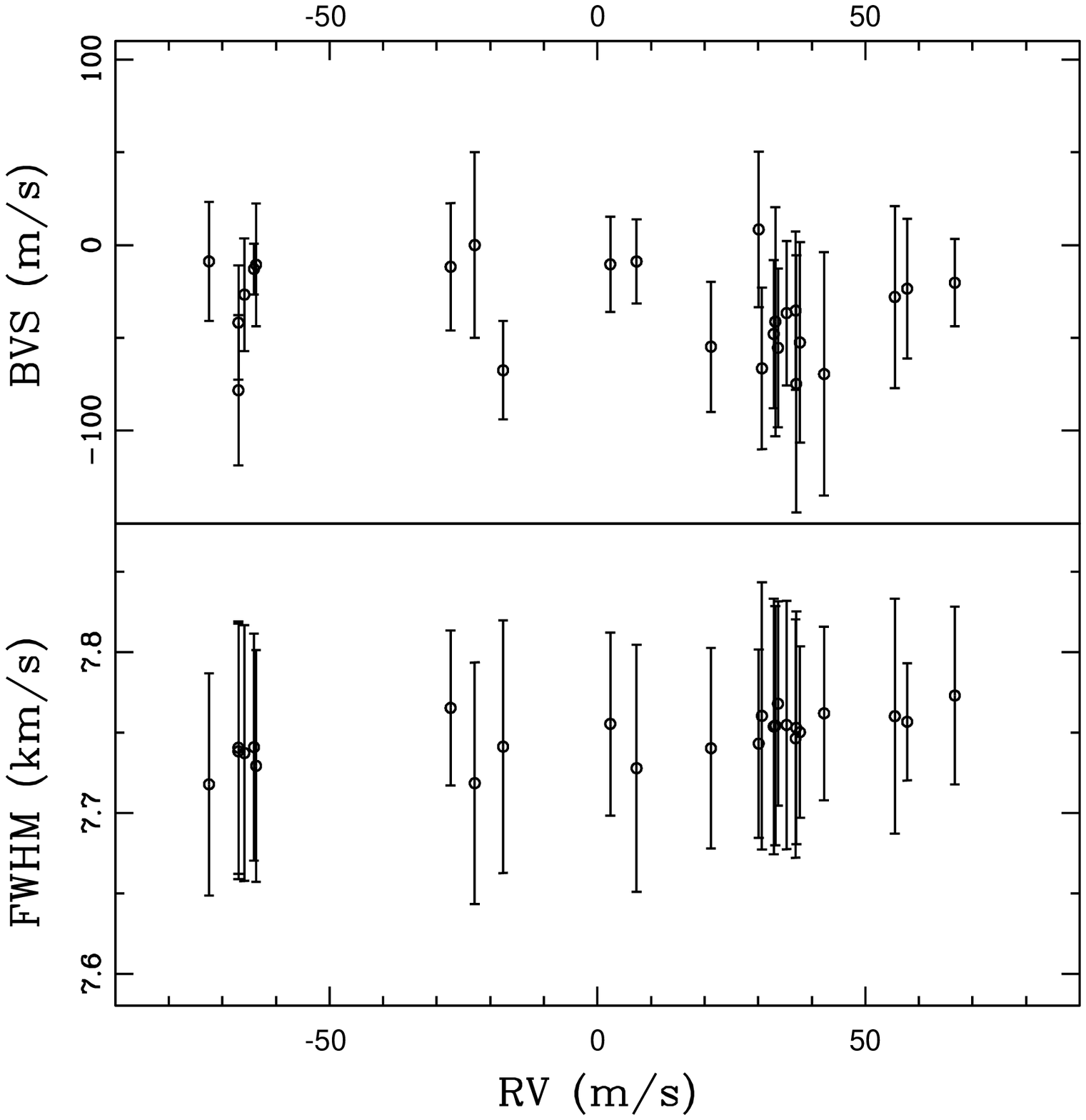}
\caption{Bisector velocity span and full width at half maximum of the CCF versus FEROS radial velocities of HIP\,67851 (upper and lower panel,
respectively).\label{HIP67851_BVS}}
%Bisector velocity span against the FEROS RVs of HIP\,67851. 
%The BVS at each epoch was computed from the mean BVS in 11 different orders. The error bars correspond to the uncertainty in the mean.
%Lower panel: Full width at half maximum of the CCF versus the HIP\,67851 radial velocity measurements.  \label{HIP67851_BVS}}
\end{figure}

\begin{figure}[h]
\centering
\includegraphics[width=7cm,height=5cm,angle=0]{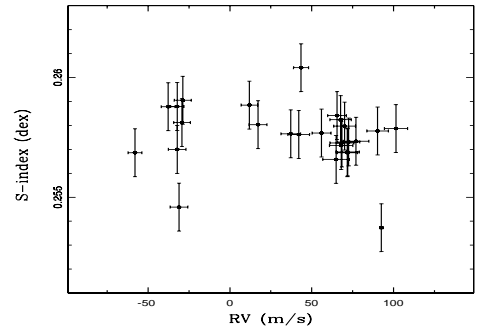}
\caption{Chromospheric activity indicator (S-index) versus the HIP\,67851 radial velocities.  \label{HIP67851_activities}}
\end{figure}

Figure \ref{HIP67851_vels} shows the RV measurements of HIP\,67851. The black dots and red triangles correspond to FEROS and CHIRON 
velocities, respectively. The RV variations are also listed in the electronic Tables \ref{feros_HIP67851_vels} and 
\ref{chiron_HIP67851_vels}.
The solid line represents one possible Keplerian solution with an outer planet with M$_P$ = 6.3 \mjup, P = 2200 days and zero eccentricity.
The orbital parameters of the inner planet (HIP\,67851\,{\it b}) are listed in Table \ref{HIP67851_orb_par}.
The orbital fit was computed using the systemic console (Meschiari et al. \cite{MES09}). The uncertainties were computed using a bootstrap randomization 
method. The RMS around the fit is 9.4 \ms. As can be seen, although the orbital parameters of the inner planets are 
well determined, further follow-up is needed to fully constrain the orbital period of the outer object, which probably has a mass in the
planetary regime. The best Keplerian solution for HIP\,67851\,{\it b} yields an orbital distance of 0.46 AU, meaning that it is the second innermost 
planet detected around any giant star.\footnote{The innermost planet detected around a giant star is Kepler 91\,{\it b} (Lillo-Box et al. 
\cite{LILLO14}; Barclay et al.  \cite{BAR14}). In addition, HIP\,13044\,{\it b} was shown to likely not be a real planet (Jones \& Jenkins \cite{JONJEN14})}  
%
% (after Kepler 91; see Lillo-Box et al. \cite{LILLO14}; Barclay et al. \cite{BARO14}). 
%\footnote{The planets in short-period orbits HIP\,13044\,{\it b} (Setiawan et al. \cite{SET10}) and 
%{\it Kepler-91\,b} (Lillo-Box et al. \cite{LILLO14}) have been shown to likely not be real planets. 
%See Jones \& Jenkins (\cite{JONJEN14}) and Sliski \& Kipping (\cite{SLI14}).} \newline \indent
We tested the nature of the periodic RV variations observed in HIP\,67851 using the Hipparcos photometry of this star. This dataset is comprised 
of 75 high-quality measurements. No significant periodic signal is observed in the data and the RMS is only 0.009 mag.
In addition, we performed a line profile variation analysis. Figure \ref{HIP67851_BVS} shows the bisector velocity span (BVS) and the full width at half maximum (FWHM)
of the cross-correlation function (CCF) versus the measured radial velocities (upper and lower panel, respectively). 
These values were computed according to the procedure explained in Jones et al. \cite{JON13}.
Clearly there is no correlation between these two quantities and the RV measurements.
Finally, Figure \ref{HIP67851_activities} shows the chromospheric activity index (S-index) as a function of the HIP\,67851 RVs. These values were computed according to 
the procedure described in Jenkins et al. (\cite{JEN08,JEN11}). No obvious correlation is observed between these quantities.
Based on the results of these three tests, along with a failed periodogram search for any frequencies present in those data, 
we conclude that the RV variations observed in HIP\,67851 are most likely induced by two substellar objects.
\onltab{
\begin{table}
\centering
\caption{FEROS radial velocity variations of HIP\,67851\label{feros_HIP67851_vels}}
\begin{tabular}{lcc}
\hline\hline
JD\,-        & RV & error  \\
2455000   &  (\ms)   &   (\ms) \\
\hline
   317.647 & -58.0 & 4.2 \\
   379.651 &  11.9 & 5.0 \\
   729.632 &  92.6 & 1.0 \\
   744.602 &  42.1 & 6.6 \\
  1047.622 &  68.5 & 7.3 \\
  1056.609 &  65.5 & 5.7 \\
  1066.625 &  90.3 & 6.6 \\
  1099.610 &  37.2 & 6.0 \\
  1110.584 &  17.2 & 5.4 \\
  1140.607 &  43.5 & 4.6 \\
  1321.799 &  64.9 & 8.1 \\
  1331.734 &  77.1 & 7.8 \\
  1331.806 &  72.6 & 6.8 \\
  1342.780 & 101.5 & 7.1 \\
  1412.574 &  56.0 & 5.8 \\
  1431.540 &  67.7 & 6.5 \\
  1431.587 &  70.1 & 6.7 \\
  1431.640 &  68.0 & 7.1 \\
  1431.686 &  71.8 & 6.2 \\
  1431.750 &  72.0 & 7.1 \\
  1472.606 & -29.3 & 5.1 \\
  1472.610 & -32.2 & 5.4 \\
  1472.627 & -28.8 & 5.2 \\
  1472.643 & -37.7 & 4.2 \\
  1472.657 & -32.2 & 4.1 \\
  1472.712 & -31.1 & 5.4 \\
\hline\hline
\end{tabular}
\end{table}
}
\onltab{
\begin{table}
\centering
\caption{CHIRON radial velocity variations of HIP\,67851\label{chiron_HIP67851_vels}}
\begin{tabular}{lcc}
\hline\hline
JD\,-        & RV & error  \\
2455000   &  (\ms)   &   (\ms) \\
\hline
  1299.843 &  7.6 &4.3 \\
  1307.835 & 26.2 &4.3 \\
  1312.800 & 32.6 &4.4 \\
  1317.784 & 37.9 &4.5 \\
  1331.897 & 78.0 &4.6 \\
  1338.805 & 93.1 &4.1 \\
  1374.783 & -8.5 &4.0 \\
  1384.685 & -7.2 &4.4 \\
  1385.645 & -1.3 &4.3 \\
  1411.529 & 28.4 &4.7 \\
  1422.692 & 67.7 &4.4 \\
  1426.595 & 67.6 &4.3 \\
  1476.654 &-32.5 &3.7 \\
  1523.469 & 38.6 &4.0 \\
  1533.483 & 23.2 &5.1 \\
  1539.480 & 13.2 &9.4 \\
  1654.819 &-82.3 &6.2 \\
  1656.815 &-71.1 &5.7 \\
  1659.806 &-59.7 &4.5 \\
  1666.858 &-37.0 &5.0 \\
  1670.816 &-12.2 &4.9 \\
  1676.760 & -5.4 &5.0 \\
  1683.887 & 13.8 &5.4 \\
  1690.891 & 20.1 &6.0 \\
  1708.828 &  3.3 &4.5 \\
  1722.691 &-47.1 &4.3 \\
  1736.616 &-92.0 &5.2 \\
  1752.679 &-70.8 &4.5 \\
  1767.540 &-43.6 &4.5 \\
  1783.751 &-17.8 &5.4 \\
  1798.616 &-22.2 &5.6 \\
  1819.563 &-94.0 &6.5 \\
  1823.577 &-86.8 &5.4 \\
  1840.551 &-86.3 &4.3 \\
\hline\hline
\end{tabular}
\end{table}
}
\begin{table}
\centering
\caption{Orbital parameters of HIP\,67851\,b and HIP\,97233\,{\it b}.\label{HIP67851_orb_par}}
\begin{tabular}{lcc}
\hline\hline
                   &   HIP\,67851\,{\it b}    & HIP\,97233\,{\it b}  \\
\hline 
P (days)           &   88.8     $\pm$  0.2    & 1058.8  $\pm$ 6.7  \\
K (\ms)            &   46.8     $\pm$  2.0    & 320.1   $\pm$ 18.4 \\
a (AU)             &   0.46     $\pm$  0.01   & 2.55    $\pm$ 0.05 \\
e                  &   0.09     $\pm$  0.05   & 0.61    $\pm$ 0.03 \\
$\omega$ (deg)     &   87.7     $\pm$  36.4   & 249.3   $\pm$ 2.4  \\
M\,sin$i$ (\mjup)  &   1.44     $\pm$  0.04   & 20.0    $\pm$ 0.4  \\
T$_P$ (JD-2455000) &   296.6    $\pm$  7.7    & 856.3   $\pm$ 10.0 \\
\hline\hline
\end{tabular}
\end{table}

\subsection{HIP\,97233\,{\it b}} 

\begin{figure*}[]
\centering
\includegraphics[width=13cm,height=10.5cm,angle=0]{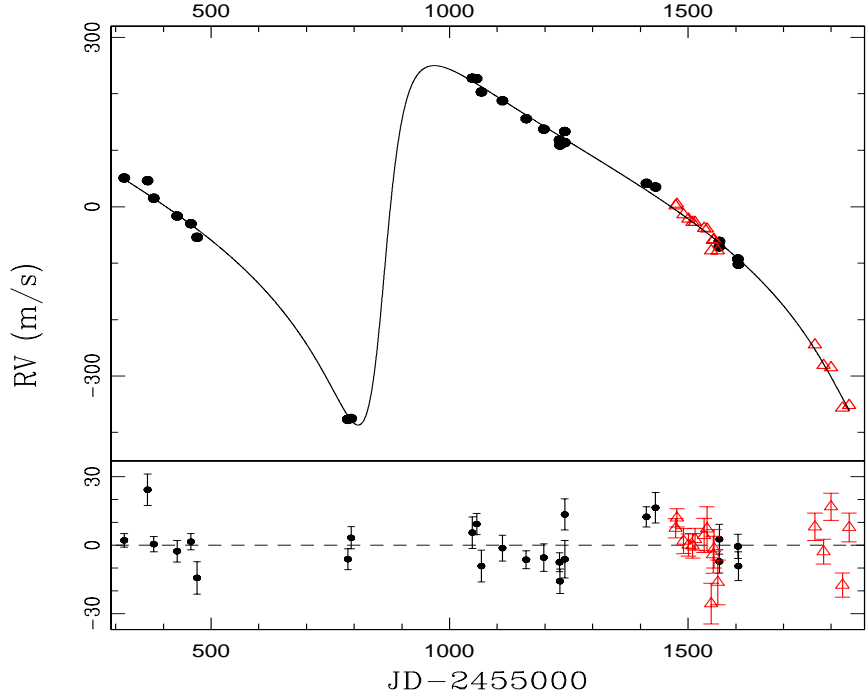}
\caption{Upper panel: Radial velocity curve for HIP\,97233. The red triangles and black dots
correspond to CHIRON and FEROS velocities, respectively. The best Keplerian solution is overplotted (black solid line).
Lower panel: Residuals from the Keplerian fit. The RMS around the fit is 10.1 \ms.  \label{HIP97233_vels}}
\end{figure*}

\begin{figure}[ht]
\centering
\includegraphics[width=7.5cm,height=9cm,angle=0]{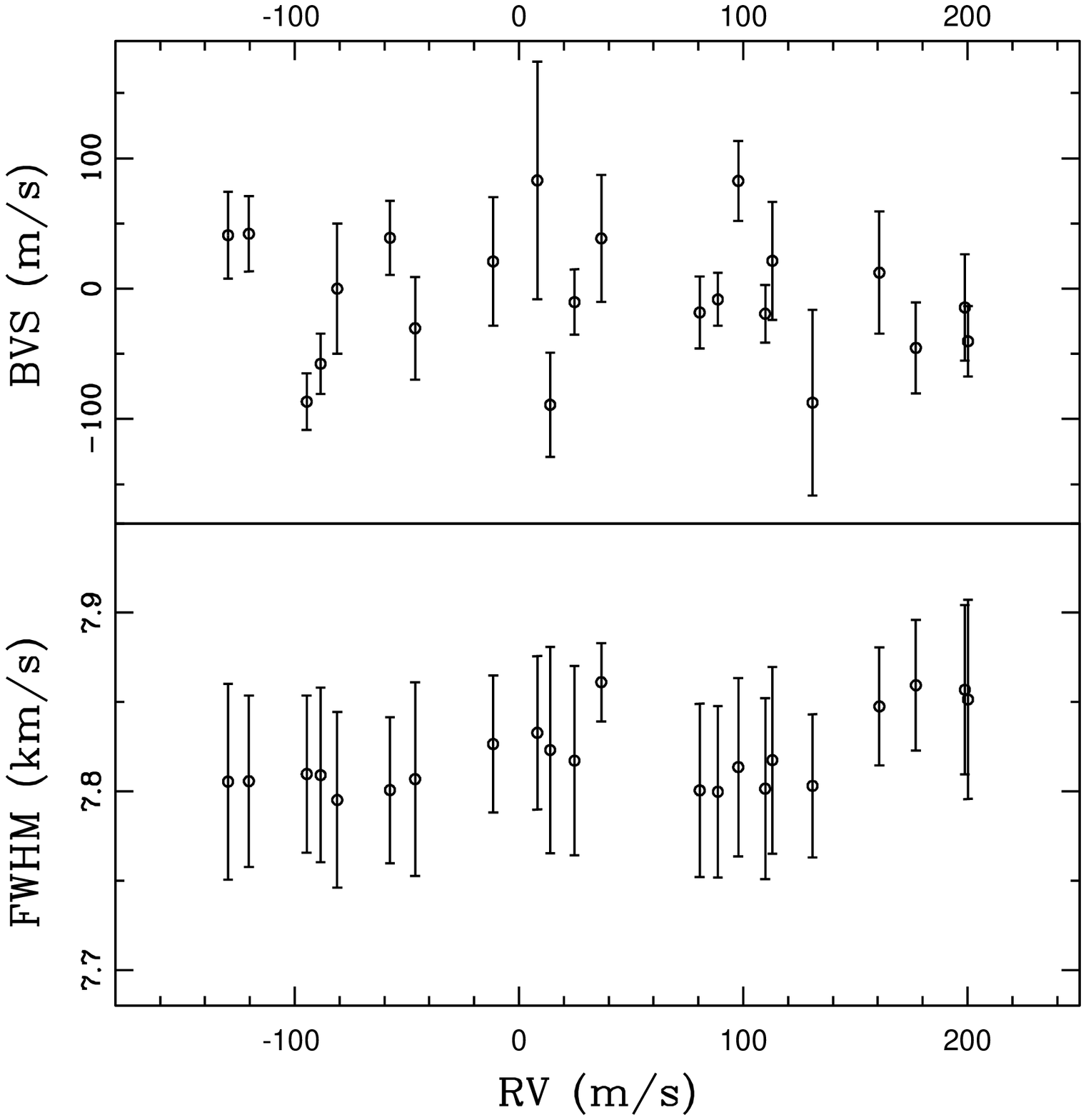}
\caption{Bisector velocity span and full width at half maximum of the CCF versus FEROS radial velocities of HIP\,97233 (upper and lower panel,             
respectively). \label{HIP97233_BVS}}
\end{figure}

\begin{figure}[]
\centering
\includegraphics[width=7cm,height=5cm,angle=0]{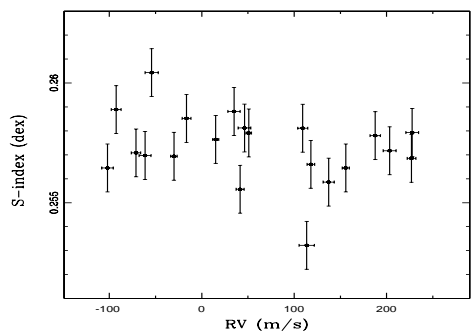}
\caption{Chromospheric activity indicator (S-index) versus the HIP\,97233 radial velocities.  \label{HIP97233_activities}}
\end{figure}

Figure \ref{HIP97233_vels} shows the RV measurements of HIP97233. The black dots and red triangles correspond to FEROS and CHIRON velocities, 
respectively. The solid line represents the best Keplerian fit. The orbital parameters are listed in Table \ref{HIP67851_orb_par}.
%The uncertainties were computed with the systemic console (Meschiari et al. \cite{MES09}) using a boostrap randomization method.
The RMS around the fit is 10.1 \ms, which is consistent with the RV noise induced by 
stellar oscillations in giant stars (e.g. Hekker et al. \cite{HEK06}).
The RV variations are also listed in Tables \ref{feros_HIP97233_vels} and \ref{chiron_HIP97233_vels}. \newline \indent
As for HIP\,67851, we investigated different possible mechanisms that might be causing the RV variations observed in HIP\,97233. 
We analyzed the Hipparcos photometric data of this star, which consists of a total of 46 high quality observations.
There is no significant periodic signal in the data and the RMS is below 0.01 mag. Such a low variability is not expected to mimic a signal as 
large as the signal observed in the HIP\,97233 data (Hatzes \cite{HAT02}). 
Additionally, we performed a line profile analysis. The results from the BVS and FWHM 
variations are plotted in Figure \ref{HIP97233_BVS} (upper and lower panel, respectively).   
%The upper and lower panel show the bisector velocity span (BVS) and the full width at half maximum (FWHM) variations of the 
%cross-correlation function (CCF), respectively. These values were computed according to the procedure explained in Jones et al. \cite{JON13}.
No obvious correlation is observed between these values and RV variations. Finally, Figure \ref{HIP97233_activities} shows the S-index versus the observed 
radial velocities of HIP\,97233. Clearly, there is no correlation with the radial velocities. 
Based on these results we conclude that the most likely explanation for the periodic RV variations observed in HIP\,97233 is due to the presence
of a substellar object. 
\onltab{
\begin{table}
\centering
\caption{FEROS radial velocity variations of HIP\,97233\label{feros_HIP97233_vels}}
\begin{tabular}{lcc}
\hline\hline
JD\,-        & RV & error  \\
2455000   &  (\ms)   &   (\ms) \\
\hline
  317.821  &   50.9 & 3.0 \\
  366.873  &   46.2 & 6.9 \\
  379.827  &   15.1 & 3.3 \\
  428.705  & -16.5  &4.7 \\
  457.619  & -30.1  &3.5 \\
  470.603  & -54.2  &7.1 \\
  786.752  &-376.7  &4.6 \\
  793.742  &-375.3  &4.9 \\
 1047.819  & 227.7  &6.9 \\
 1056.778  & 226.9  &4.7 \\
 1066.788  & 203.3  &6.9 \\
 1110.761  & 187.7  &5.6 \\
 1160.694  & 155.9  &3.9 \\
 1197.626  & 137.3  &6.0 \\
 1230.525  & 118.0  &4.1 \\
 1231.500  & 109.2  &5.4 \\
 1241.487  & 133.2  &6.8 \\
 1241.492  & 113.6  &8.2 \\
 1412.790  &  41.5  &4.4 \\
 1431.767  &  34.9  &6.7 \\
 1565.547  & -71.1  &5.1 \\
 1565.596  & -61.4  &6.5 \\
 1604.491  & -92.6  &5.3 \\
 1605.503  &-102.0  &6.3 \\
\hline\hline
\end{tabular}
\end{table}
}
\onltab{
\begin{table}
\centering
\caption{CHIRON radial velocity variations of HIP\,97233\label{chiron_HIP97233_vels}}
\begin{tabular}{lcc}
\hline\hline
JD\,-        & RV & error  \\
2455000   &  (\ms)   &   (\ms) \\
\hline
 1473.851 &   1.7  &  4.2 \\
 1476.798 &   4.5  &  4.0 \\
 1490.792 &  -14.0 &  5.5 \\
 1500.740 &  -21.8 &  4.8 \\
 1509.719 &  -27.7 &  5.4 \\
 1514.623 &  -27.9 &  4.8 \\
 1533.651 &  -38.3 &  7.5 \\
 1539.628 &  -39.3 &  9.6 \\
 1548.606 &  -78.0 &  8.9 \\
 1552.511 &  -59.2 &  8.1 \\
 1554.550 &  -58.0 &  8.4 \\
 1562.531 &  -78.1 &  9.8 \\
 1765.847 & -244.7 &  6.0 \\
 1783.910 & -280.8 &  5.4 \\
 1799.840 & -284.9 &  5.9 \\
 1823.733 & -356.4 &  5.3 \\
 1837.707 & -352.2 &  6.3 \\
\hline\hline
\end{tabular}
\end{table}
}
\section{Summary and discussion.}

In this paper we present the discovery of two substellar systems found around the giant stars HIP\,67851 and HIP\,97233.
The radial velocity variations of HIP\,67851 are most likely explained by the presence of two planetary-mass objects. 
The inner planet in the system has the following orbital parameters:
M$_{\rm P}\,$sin$i$ = 1.44 \mjup, P = 88.8 days and $e$ = 0.09. HIP\,67851\,{\it b} is one of the few planets orbiting interior to $\sim$ 0.6 AU 
around giant stars.
The orbital properties of the outer object are still uncertain, since only a fraction of its orbital 
period has been covered. However, it is probably a super-Jupiter or a BD in a wide orbit ($a \gtrsim$ 4 AU). %\newline \indent 
This new discovery confirms the finding by Schlaufman \& Winn (\cite{SCH13}) and Jones et al. (\cite{JON14}) since HIP67851\,{\it b} has a nearly
circular orbit ($e \lesssim$ 0.2) and a semi-major axis $\lesssim$ 0.9 AU. 
Moreover, if HIP67851\,c is actually a super-Jupiter, it would also confirm that most of the planets around giant stars are super-planets. 
\newline \indent 
On the other hand, the RV curve of HIP\,97233 shows a periodic signal that is most likely induced by a BD with the following orbital
parameters: M$_{\rm P}\,$sin$i$ = 20.0 \mjup, P = 1058.8 days and $e$ = 0.61. 
With this new discovery there are $\sim$ 10 known BDs around giant stars, with HIP\,97233\,{\it b} being the most eccentric of these.
Although this number is small, it is similar to the number of known BDs orbiting interior to $\sim$ 4 AU around solar-type stars.
However, since the number of giant stars targeted by RV surveys is relatively small, %a fraction of the solar-type star sample targeted by these surveys,
it seems evident that the fraction of BDs around giant stars is significantly higher than around solar-type stars (Mitchell et al. \cite{MIT13}).   
The reason for this difference in the detection rate is not due to an observational bias, rather it is probably explained by either the role of the 
host star mass (since the giant host stars are mainly intermediate-mass stars) or by the effect of the stellar evolution. 
In the former case, the mass and content of the protoplanetary disc might strongly affect the efficiency of planet formation 
and the subsequent gas accretion. Thus, it is possible that intermediate-mass stars have more massive and dense discs, from which massive planets
and BD are more efficiently formed (Lovis \& Mayor \cite{LOV07}). On the other hand, the mass of these super-planets might be explained by the 
evolution of the host star. In this scenario, a giant planet might be able to accrete a 
significant amount of mass from an enhanced stellar wind during the giant branch phase, eventually exceeding the deuterium burning 
limit and becoming a BD (Jones et al. \cite{JON14}).  
It is also notable that the metallicity of HIP\,97233 is very high ([Fe/H]\,=\,0.29\,$\pm$\,0.13~dex). 
It might thus be possible that its brown dwarf companion might have formed through core accretion (see Jenkins et al. \cite{JEN09}), 
in a process similar to what is thought to be the dominant formation mechanism for planets.
% and this could also lead to the development of a massive body orbiting this star through the core accretion planet formation process (see Jenkins et al. 
%\cite{JEN09}), a finding that would suggest these planets/BDs could have formed in a similar manner to planets orbiting lower-mass stars.  
If this is the case, then 
we would expect to detect more massive objects orbiting the most metal-rich stars since these discs form higher-mass cores more efficiently, giving rise 
to a deficit of low-mass objects orbiting the most metal-rich stars (Jenkins et al. \cite{JEN13}).  
Finally, due to the interaction with either the disc before
the gas is dissipated or with the stellar envelope during the post-MS phase, these BDs might have migrated from beyond $\sim$ 4 AU to their 
current position. In this scenario the primordial fraction of BDs could be the same as for solar-type host stars.

\begin{acknowledgements}
M.J. acknowledges financial support from Fondecyt project \#3140607.
J.J and P.R. acknowledge funding from CATA (PB06, Conicyt). P.R. acknowledges support from
Fondecyt project \#1120299. We acknowledge the anonymous referee for useful comments.
This research has made use of the SIMBAD database and the VizieR catalogue access tool, operated at CDS, Strasbourg, France. 
%P.R. acknowledges financial support from Fondecyt through grant \#1120299 and Conicyt-PIA Anillo ACT1120.
%J.J. acknowledges funding by Fondecyt through grant 3110004, 
%the GEMINI-CONICYT FUND and from the Comit\'e Mixto ESO-GOBIERNO DE CHILE. 
%M.J., J.J, and P.R also acknowledge support from BASAL PFB-06 (CATA).
\end{acknowledgements}

\end{document}